\documentclass[aps,twocolumn,showpacs]{revtex4}

\usepackage{graphicx}
\usepackage{amssymb}

\def\ltab#1{{\tabcolsep=0pt\begin{tabular}{l}#1\end{tabular}}}

\begin{document}
\title{Proton ejection from molecular hydride clusters exposed to strong X-ray pulses}

\author{Pierfrancesco Di\,Cintio, Ulf Saalmann, and Jan-Michael Rost}
\affiliation{Max Planck Institute for the Physics of Complex Systems\\
  N\"othnitzer Stra{\ss}e 38, 01187 Dresden, Germany}
\date{\today}

\begin{abstract}\noindent 
Clusters consisting of small molecules containing hydrogen do eject fast protons when illuminated by short X-ray pulses.  
A suitable overall charging of the cluster controlled by the X-ray intensity induces electron migration from the surface to the bulk
leading to efficient segregation of the protons and to a globally hindered explosion of the heavy atoms even outside the screened volume. We investigate this peculiar effect systematically along the iso-electronic sequence of methane over ammonia and water to the atomic limit of neon as a reference.  In contrast to core-shell systems where the outer shell is sacrificed to reduce radiation damage, the intricate proton dynamics of hydride clusters allows one to keep  the entire backbone of heavy atoms intact.
\end{abstract}

\pacs{36.40.-c, 32.80.Aa, 33.80.Eh, 79.77.+g} 
\maketitle

\noindent
Composite clusters consisting of different atomic or molecular species can have surprising dynamical properties under intense light pulses.
This has been demonstrated for so called core-shell systems where the core-cluster is formed by one sort of atoms and the hull by another one \cite{hobo+08,zich+11,suiw+12}. A quite spectacular effect was predicted \cite{misa+09} and experimentally verified \cite{krfe+11} for helium-embedded rare-gas clusters which absorb  near-infrared photons extremely efficiently \cite{misa+09}. 
For this to happen the core material must have a lower ionization potential than the hull which together, with a spatially preferred region of ionization  due to the polarization of the light generates an anisotropic electron plasma which can be resonant with the laser frequency for a long time. Seed atoms with a lower ionization potential than the rest of the material are probably also responsible for a dramatic enhancement of light absorption in rare-gas clusters ``contaminated'' by a very small fraction (of a few percent) of water \cite{jhkr08}. 

Here, we investigate a completely different kind of composite clusters, namely ``heavy-light'' systems composed of hydride molecules. To be specific, we will concentrate on the iso-electronic sequence of methane (CH$_{4}$), ammonia
(NH$_{3}$) and water (H$_{2}$O) clusters, augmented by the ``atomic limit'' of neon clusters.  Apart from being genuinely interesting in the context of composite clusters, these systems contain the  elements hydrogen, carbon, nitrogen and oxygen, omnipresent  in organic molecules. A detailed understanding of their dynamics in strong X-ray pulses provides valuable information to realize coherent diffractive imaging with single molecules \cite{spwe+12} in the future.

Protonated large systems can eject fast protons upon energy absorption from X-ray pulses \cite{beti+04,iwan+12}. As we will demonstrate here, the proton loss provides an effective additional channel to release the absorbed energy and acquired charge as compared to systems which do not contain protons with intriguing consequences: 
(i) The  nano-plasma from trapped electrons is much cooler than in pristine clusters without hydrogen but the same number of heavy atoms. 
(ii) The proton ejection dynamics exhibits universal features along the iso-electronic hydrides but very different from the iso-electronic atomic cluster. 
(iii) The probably most surprising effect is a \emph{dynamically induced segregation\/} of heavy ions and protons for which field ionization \cite{gnsa+09} plays a prominent role. 
It is very different from simplified models for multi-component Coulomb explosion \cite{anni+10}. 
The dynamical segregation occurs in an experimentally relevant intensity window  $I\,{\sim}\,10^{17}\ldots10^{18}$\,W/cm$^{2}$ where the heavy atoms (C, N or O) emerge as \emph{neutrals\/} despite substantial energy deposition by the laser pulse. This is in sharp contrast to pristine clusters composed out of C, N$_{2}$ or O$_{2}$ which Coulomb explode with multiply charged ions for a comparable total charging. 

Our theoretical description relies on a mixed quantum-classical approach. Electrons and ions are treated as classical particles and are propagated according to Newton's equations with all Coulomb interactions included \footnote{In order to prevent unphysical auto-ionization in the classical calculations we use a smoothed Coulomb interaction $W(r)=1/\sqrt{r^{2}+1}$ between particles with a distance $r$.}. Photo-ionization and Auger decay, i.e., the quantum electronic processes within the atoms or hydrides, are described by the appropriate rates and consistently integrated into the time-evolution of the charged particles through a Monte-Carlo realization \cite{gnsa+09,gesa+07gnsa+12}.
In the present study, we use X-ray pulses with a photon energy of $\hbar\omega=1{\rm keV}$ and a Gaussian temporal envelope with full-width-at-half-maximum of $T=10$\,fs and peak intensities $I=10^{16}{\ldots}10^{19}{\rm W/cm^2}$. 
The pulse lasts longer than Auger decays (cf.\ Table~\ref{tab:data}) but is still shorter than typical expansion times of the charged cluster. 

\begin{table}[bh]
\caption{$K$-shell parameters for the respective heavy atom used in the calculations.  All values refer to neutral systems.}
\tabcolsep=4pt
\begin{tabular}{|lc|c|c|c|c|}\hline
&& Ne & H$_{2}$O & NH$_{3}$ & CH$_{4}$ \\ \hline
 binding energy $\varepsilon$ in eV &\cite{thva09}&  870.2 & 543.1 & 409.9 & 284.2 \\ 
\ltab{photo-ionization  (1\,keV)\\ [-0.7ex]  cross section $\sigma$ in kb} & \cite{behu+98} 
& 248.2 & 121.9 & 76.99 &     44.07 \\ 
Auger life-time $\tau$ in fs &\cite{koav11}& 2.9 & 4.46 & 5.36 &  7.76 \\ 
\hline
\end{tabular}
\label{tab:data}
\end{table}%

The probability for  a photo-ionization event  to happen between  $t$ and $t{+}\Delta t$ for the $j$-th shell of the $i$-th atom or molecule in the cluster is given by
\begin{equation}\label{probphotoion}
{\mathcal P}^{{\rm photo}}_{j,i}=I_t\frac{\sigma_{j,i}\Delta t}{\hbar\omega};\quad j=\mbox{1s,\,2s,\,2p},
\end{equation}
where $I_t=I\exp(-4\ln2(t/T)^2)$ is the laser intensity at time $t$ and $\sigma_{j,i}$  photo-ionization cross section for the neutral species with $n_{\rm e}$ active electrons listed in Table~\ref{tab:data}. If the $j$th shell has $n_{\rm v}$ vacancies, ${\mathcal P}^{{\rm photo}}_{j,i}$ is reduced by a factor $\chi=(n_{\rm e}-n_{\rm v})/n_{\rm e}$.
Photons of a few-keV energy ionize mainly the $K$-shell (1s-orbitals)  for the elements of the first row under consideration. 
In our test simulations for CH$_{4}$ less than 3\,\% of the photo-electrons come from the valence shell at peak intensities $I\,{\approx}\,10^{19}~{\rm W/cm^2}$. Hence, for the sake of simplicity, all results presented have been obtained with photo-ionizing exclusively $K$-shell electrons.

\begin{figure}[t]
\begin{center}
\includegraphics[width=0.8\columnwidth]{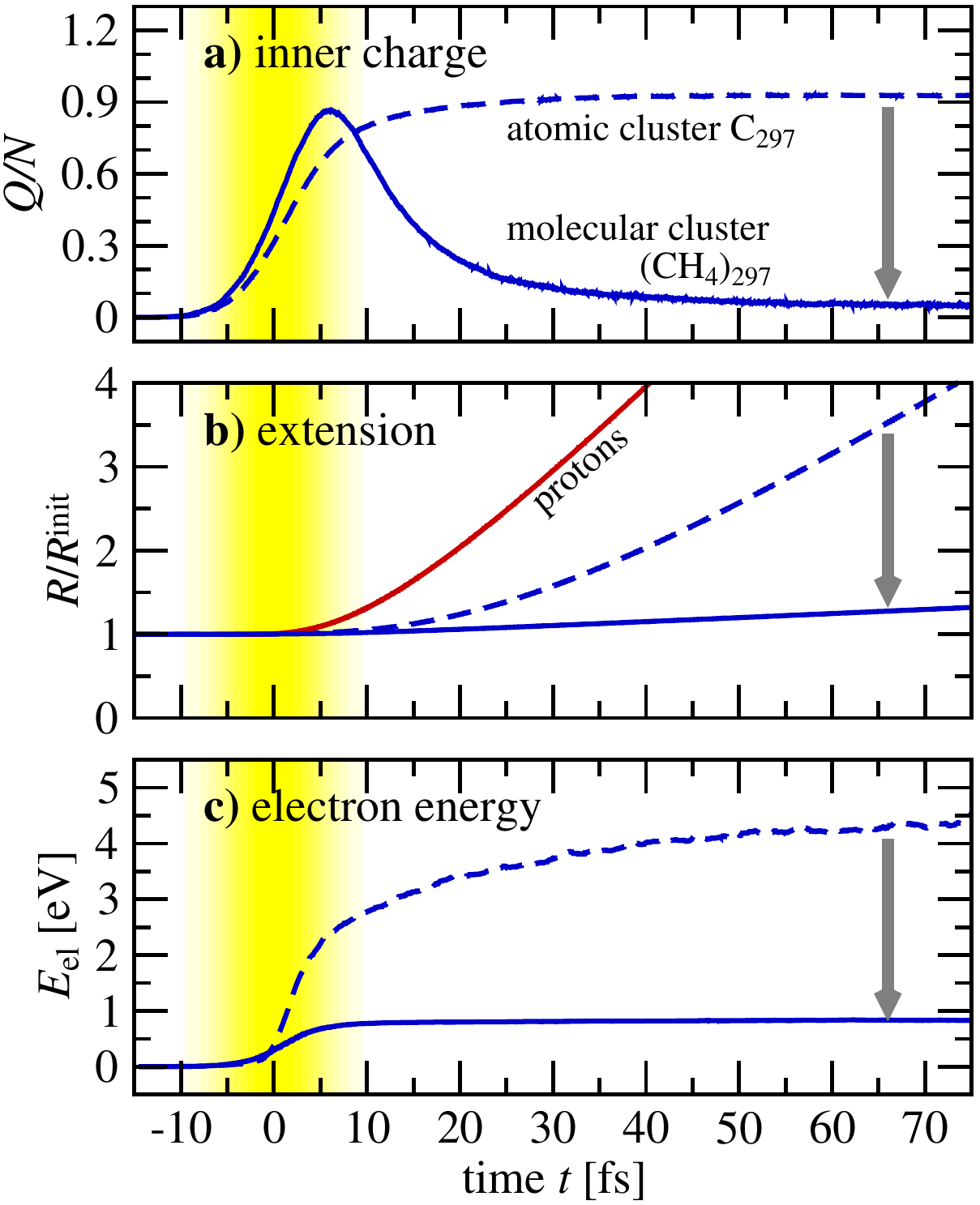}
\end{center}
\caption{Color online. Time evolution for a molecular (solid lines) and an atomic (dashed lines) cluster.
The yellow-shaded region marks the X-ray pulse ($I{=}10^{18}$W/cm$^{2}$) with a duration of 10\,fs (FWHM) centered at $t=0$.
{\bf a)} Charge inside the sphere defined by the outermost carbon ion.
{\bf b)} Radii in units of the initial radii, with the proton part shown separately (red solid line).
{\bf c)} Average kinetic energy of trapped electrons (temperature of the trapped plasma). 
The reduction of carbon charging, carbon Coulomb explosion and energy of trapped electrons due to proton ejection is marked by grey arrows.}
\label{fig:time}
\end{figure}%

$K$-shell photo-ionization produces core-hole states which decay via Auger processes \cite{me85a} with rates which may be amended by a charged environment \cite{avsa+12}. 
For a given ion $i$, the transition probability from the state $a$ to the state $b$ via an Auger decay between time  $t$ and $t{+}\Delta t$ is computed from the transition rate between the two configurations $\Gamma^{ab}_i/\hbar$ according to Fermi's golden rule weighted with the appropriate number of available transition partners,
\begin{equation}\label{probauger}
{\mathcal P}^{{\rm auger}}_{ab,i}=\frac{\Gamma^{ab}_i}{\hbar}\frac{n_{\rm e}(n_{\rm e}-1)}{n (n-1)}\Delta t,
\end{equation}
where $n_{\rm e}$ is the total number of electrons  able to take part in the transition, and $n$ is the number of electrons occupying  valence shells in the equivalent neutral species.
A similar approach for the calculation of Auger transition rates has been independently developed  \cite{inbu+12ingr+13} to describe water and methane molecules in intense X-ray lasers.

\begin{figure}[b]
\begin{center}
\includegraphics[width=0.8\columnwidth]{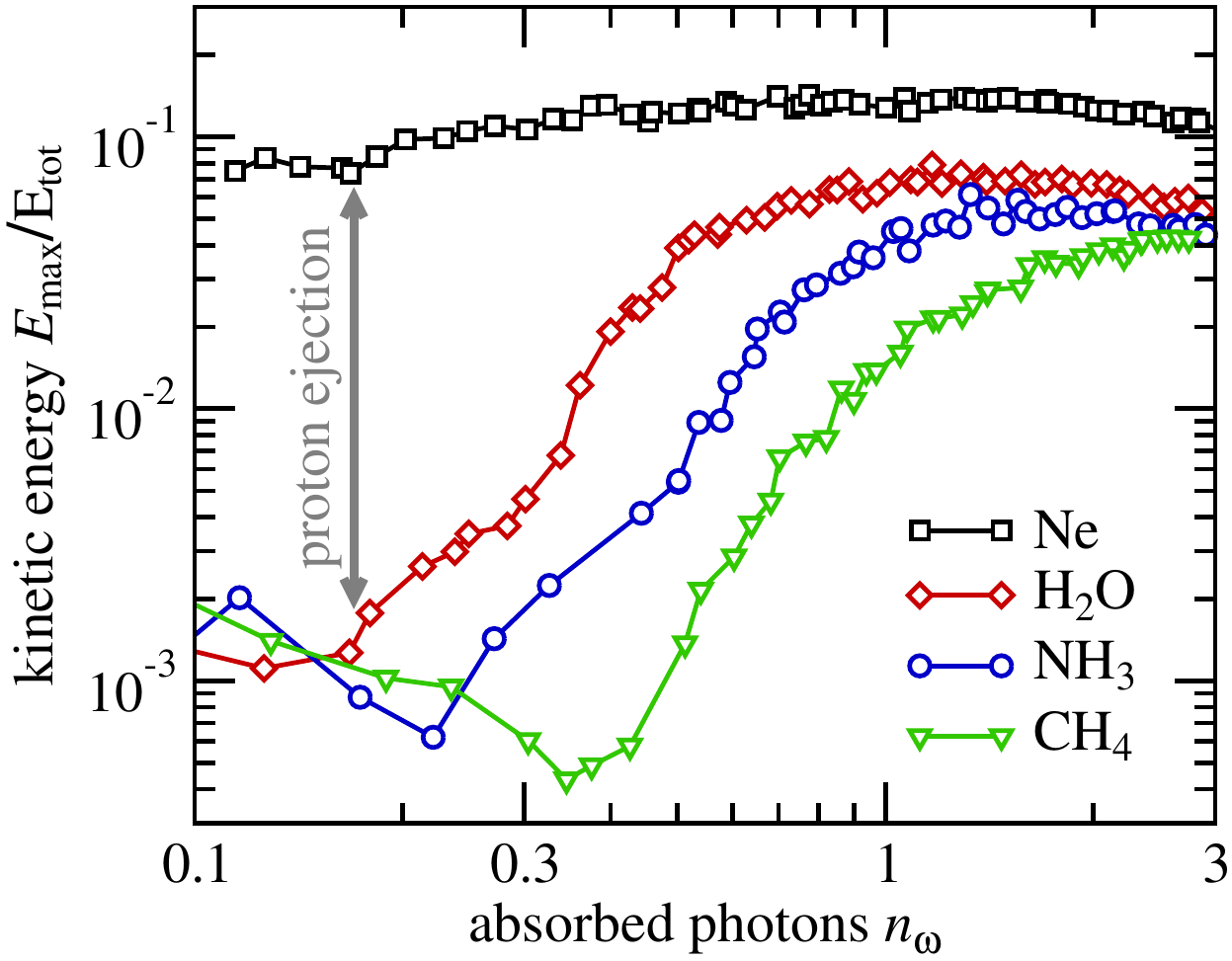}
\end{center}
\caption{Color online. Kinetic energy of the fastest ion  $E_{\rm max}$,  0.5\,ps after the peak of the pulse ($T\,{=}\,10$\,fs), versus the average number of photons absorbed per atom/molecule $n_{\omega}=N_{\omega}/N$. Cluster size is $N\,{=}\,689$. The kinetic energies are normalized with the total energy $E_{\rm tot}=N_\omega\,{\times}\,\hbar\omega$ absorbed by the cluster.}
\label{fig:Emax}
\end{figure}%
Figure \ref{fig:time} illustrates the time evolution of characteristic parameters for an atomic cluster C$_{297}$ and a molecular cluster (CH$_{4}$)$_{297}$ under the influence of the laser pulse (yellow-shaded area). One sees immediately, that the dynamics of the pristine carbon cluster (dashed) and the methane cluster (solid) is completely different, where the difference in the observables is marked  by grey arrows (at $t\,{\approx}\,65$\,fs). The carbon atoms get successively charged through photo-ionization leading to more than 90\,\% carbon ions with the laser pulse of peak intensity $I{=}10^{18}$W/cm$^{2}$ (Fig.\,\ref{fig:time}a). The cluster ions create a deep binding potential from which most Auger electrons cannot escape but form a nano-plasma. The maximum kinetic energy of the trapped electrons is limited by the depth of the cluster potential and the average kinetic energy (Fig.\,\ref{fig:time}c, dashed) is indicative of the nano-plasma temperature.
This is the normal behavior as well known from rare-gas clusters exposed to X-ray pulses \cite{sasi+06,gnsa+11}. 

The molecular cluster, however, does not follow this scheme: While initially similarly charged as in the pristine cluster, electrons recombine with the carbon ions in the methane cluster and carbon is  in the end almost neutral on average (Fig.\,\ref{fig:time}a, solid). 
At the same time the kinetic energy of the trapped electrons and hence the temperature of the nano-plasma remains comparatively low (Fig.\,\ref{fig:time}c, solid). Both phenomena originate in the ejection of fast protons from the molecular cluster
(see red line in Fig.\,\ref{fig:time}b).

Although the carbon $K$-shells are initially photo-ionized, the charge distribution 
of the doubly-charged methane after Auger decay is such that the carbon ion is screened and the positive charge is dominantly localized on two hydrogen atoms which are likely to be ejected from the entire cluster as protons. These protons take away the excess positive charge  created by photo-ionization which is of course not possible in the pristine carbon cluster. The remaining positive charge in the cluster is small giving rise to a weak potential which can only trap low-energy electrons. Therefore, the temperature of the nano-plasma is by a factor of four smaller  than for the pure carbon cluster after $t\,{\approx}\,65$\,fs (grey arrow); this holds true also for the Coulomb explosion of the carbon ions (Fig.\,\ref{fig:time}b) with the charging of the carbon ions even more dramatically reduced, almost by an order of magnitude (Fig.\,\ref{fig:time}a).

One may expect that the absolute difference in velocity of heavy and light ions gets  larger with increasing intensity and therefore higher charging of the cluster. Figure \ref{fig:Emax}, however, reveals that the ratio of the kinetic energy for the fastest heavy ion $E_\mathrm{max}$, relative to the energy of   all ions $E_\mathrm{tot}$, exhibits a non-monotonic behavior as a function of photons absorbed $n_{\omega}$ with a dip at a critical number $n_\omega^*$.  The latter depends moderately on the species considered,
as can be seen in Fig.\,\ref{fig:Emax}, but is otherwise a universal feature of hydride clusters in obvious contrast to the iso-electronic neon cluster.

\begin{figure}[t]
\begin{center}
\includegraphics[width=0.8\columnwidth]{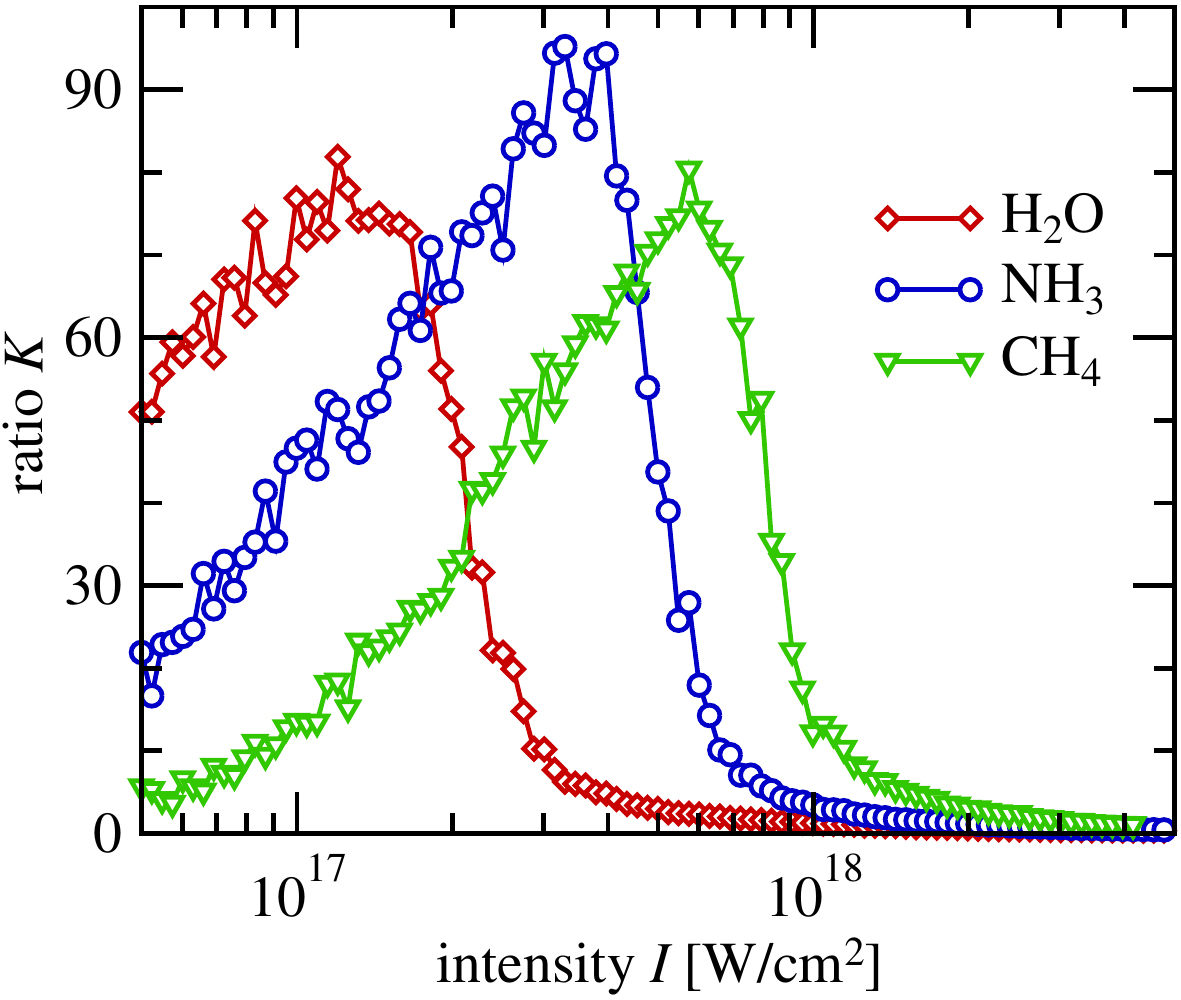}
\end{center}
\caption{Color online. Ratio $K$ of the average kinetic energy of protons and heavy atoms according to Eq.\,(\ref{eq:ratio}) for $X$=O, N, C as a function of the X-ray intensity $I$. Same parameters as in Fig.\,\ref{fig:Emax}.}
\label{fig:Eave}
\end{figure}%
The dip in the curves of Fig.\,\ref{fig:Emax} for the hydrides clusters indicates a dynamical segregation of protons and heavy ions as can be seen in Fig.\,\ref{fig:Eave}, where a more global quantity, namely the ratio of the average energy of all protons versus that of all heavy ions ($X$\,=\,O, N, C)
\begin{equation}\label{eq:ratio}
K\equiv\left<E_{\rm kin}\right>_{\rm H}\big/\left<E_{\rm kin}\right>_{\rm X},
\end{equation}
is shown as a function of peak laser intensity $I$. Again, one sees a qualitatively identical behavior of all three hydrides with a  maximal segregation at a well defined intensity although the three hydrides differ in their respective ionization energy for the 1s electrons, Auger rates and photo-ionization cross sections listed in Table \ref{tab:data}. 
Hereby the shift of the peak positions is to a large extent due to the different photo-ionization cross sections.

\begin{figure}[b]
\begin{center}
\includegraphics[width=0.8\columnwidth]{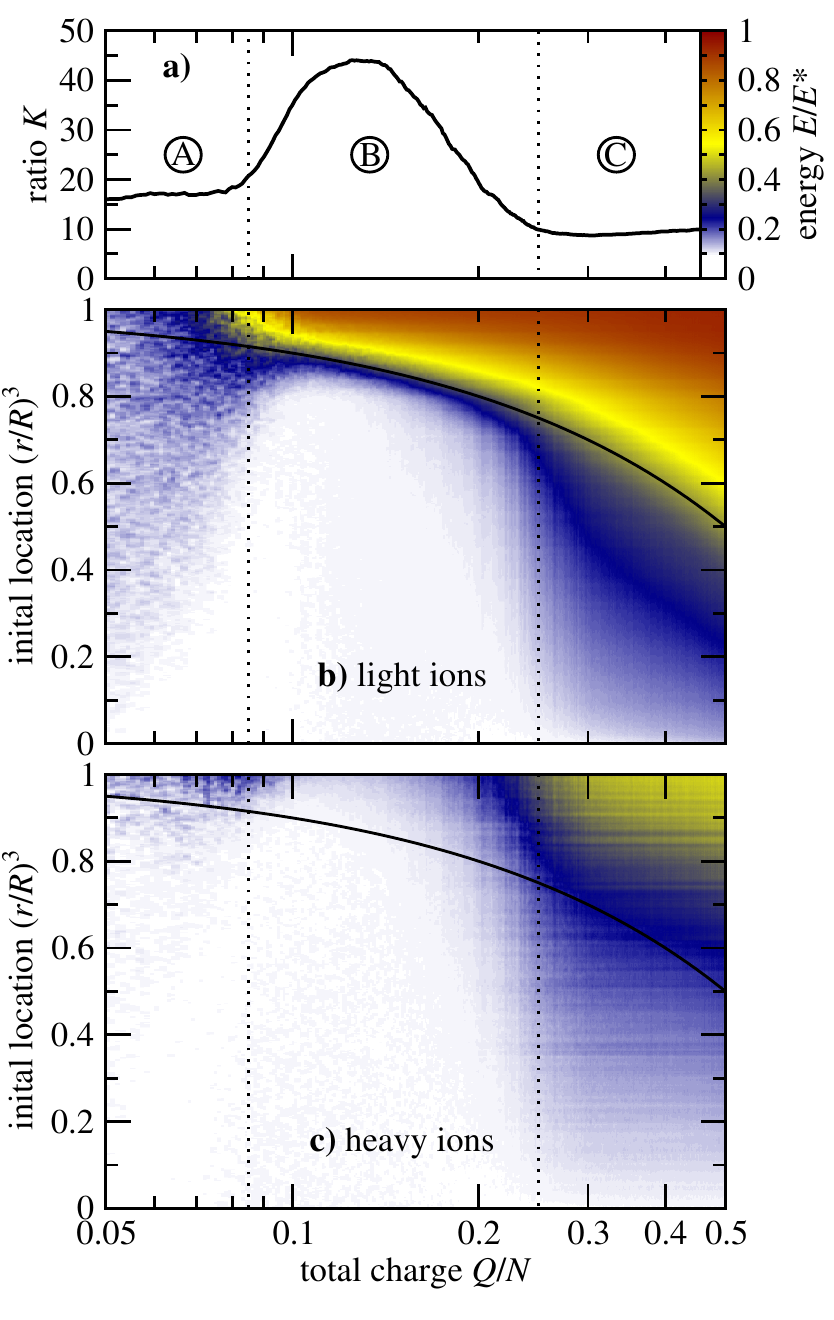}
\end{center}
\caption{Color online. Ion kinetic energies of a heterogeneous cluster model with $N{=
}10^{4}$ particles in a sphere of radius $R$ as a function of the total charge $Q/N$ of the system, with scenarios A, B, and C as discussed in the text.
\textbf{a}) Ratio of average energies of heavy and light ions, cf.\ Eq.\,(\ref{eq:ratio}) and Fig.\,\ref{fig:Eave}.
Energies resolved for the initial position  of the light (\textbf{b}) and heavy  (\textbf{c}) ions, where $(r/R)^{3}$ measures the distance from the cluster center. 
Energies are given in terms of $E^{\star}{=}Q/R$ for the light and $E^{\star}{=}Q/4R$ for the heavy component, respectively.
The solids lines indicate the part which is expected to explode if the charge would be concentrated in a shell at the surface.
}
\label{fig:model}
\end{figure}%
While the  total charge yields and particle-averaged kinetic energies have revealed the proton segregation, more differential analysis is needed to clarify wether this segregation is a local effect due to the heavy-light character of the hydride molecules or wether the segregation is a consequence of the cluster nature of the entire system. To this end we have set up a simple model where $N=10^4$ singly-charged ions are distributed homogeneously  in a sphere neither supporting any molecular sub-structure nor allowing any intra-atomic/molecular electronic processes. 
3/4 of the ions have the mass of the proton, while 1/4 has a 20 times higher mass.
$N{-}Q$ electrons are placed at randomly selected ions. With this initial configuration ions and electrons --  interacting via smoothed Coulomb forces -- are propagated for 1\,ps. Due to the positive excess charge the system fragments. 
The ratio of the average final energies of light and heavy ions, cf.\ Eq.~(\ref{eq:ratio}),
exhibits one central maximum (region B in Fig.\,\ref{fig:model}a)
similarly as for the fully microscopic calculations for hydride clusters in Fig.\,\ref{fig:Eave}. 

It is the dependence of the final energy of an ion on its initial (radial) position $r$ in the cluster (shown in Fig.\,\ref{fig:model}b and \ref{fig:model}c) which reveals the mechanism behind the heavy-light ion segregation.  In region A as well as in region C protons and  heavy ions originate from all initial positions in the cluster with an increasing energy towards the surface. The mean kinetic energy is larger in C than in A due to the stronger charging, but 
heavy-light segregation does not take place in either of the two regions. In region B, however,  the charging $Q$ is sufficiently strong to trigger field ionization of surface ions, as it has been discovered for homogeneous clusters \cite{gnsa+09}. 
As a consequence, one expects the cluster core to be screened by the field ionized electrons up to the radius indicated by a black line in Fig.\,\ref{fig:model}b,\,c. However, the protons in the hydrides cluster core are light enough (or more precise: have a sufficiently large charge-to-mass ratio) to have started moving before the repelling forces are compensated by the screening electrons. Hence, protons leave the cluster core and as a result the surplus of screening electrons prevents heavy ions in the surface layer even beyond the screening radius of a homogeneous cluster from exploding. In contrast, protons escape with high final energy from the surface layer. 

\begin{figure}[t!]
\begin{center}
\includegraphics[width=\columnwidth]{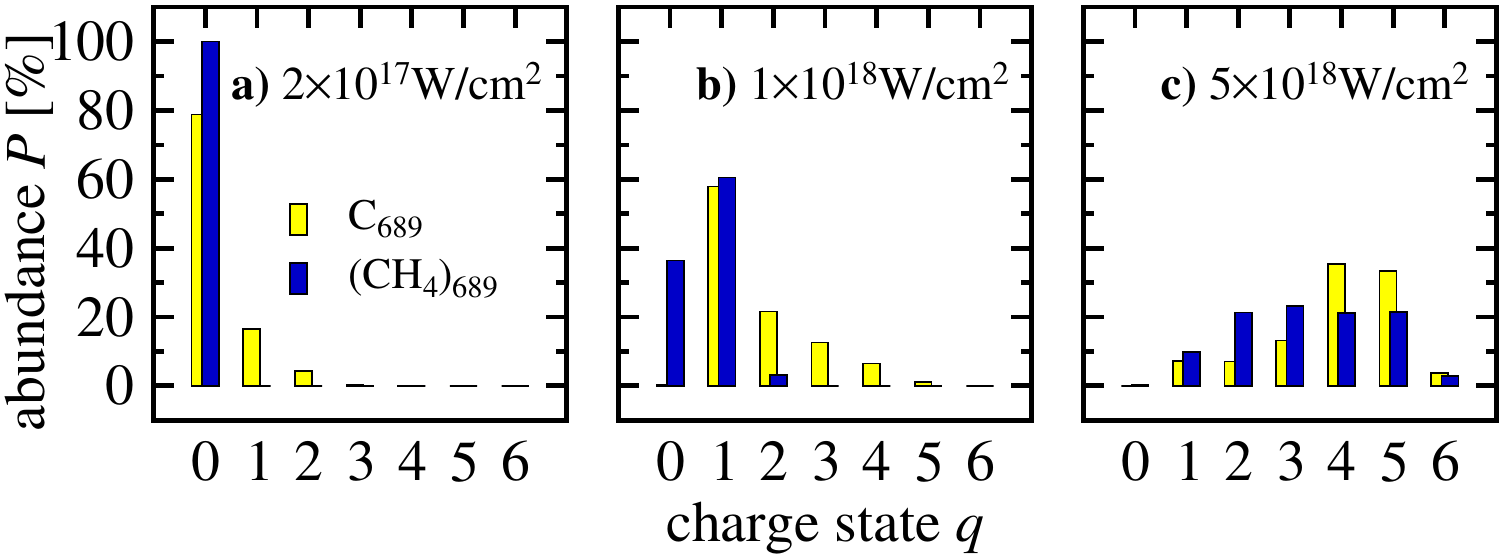}
\end{center}
\caption{Color online. Charge-state distribution for carbon ions C$^{q+}$ from (CH$_{4}$)$_{689}$ and C$_{689}$ for various X-ray intensities $I$ indicated in the respective panel.}
\label{fig:averaged-charges}
\end{figure}%
In region C the initial charge $Q$ further increases which weakens the screening effect  through field ionized electrons for two reasons. Firstly, the fraction of screening electrons available versus initial charge $Q$ decreases. Secondly,  the temperature of the screening electrons is higher than in B due to the deepening of the trapping potential with increasing $Q$. Hence, the surface layer does no longer form efficiently and a Coulomb explosion as in region A, although more violently, results.
We may conclude that heavy-light segregation is not a local effect of  heavy-light molecules. Rather, it happens in a surface layer of the heterogeneous cluster triggered by field ionization.

Having established the phenomenon of proton segregation in hydride clusters, its origin and its universal features theoretically, the question remains which kind of observable consequences the segregation has. 
We do expect a much lower charging of heavy ions as compared to the pristine cluster of the heavy-atom species (C, N or O).  This is indeed the case as can be seen in Fig.\,\ref{fig:averaged-charges}. For low intensities most carbon atoms remain neutral in the pristine as well as in the hydride cluster.
This changes drastically for intermediate intensities of about 10$^{18}$W/cm$^{2}$, where
the fraction of neutral carbon atoms surviving the light pulse illumination is small in the pristine cluster. In the  hydride cluster, one the other hand, about 80\,\% neutral heavy atoms result from recombination with the cold electrons after  proton segregation in the surface layer which has been fully charged due to efficient field ionization.
For higher intensities, we expect the proton segregation to cease (see Fig.\,\ref{fig:Emax}) and as a consequence  similar charge spectra for the pristine and the hydride cluster. This is  indeed true with respect to a vanishing yield of neutral atoms. The form of the charge distribution is still somewhat different.

What clearly emerges from Figs.\,\ref{fig:Emax} and \ref{fig:Eave} is the sensitivity of the proton ejection and consequently the charge distribution of the heavy ions on the intensity.  
Figure~\ref{fig:averaged-charges} demonstrates that significantly lower charging of the carbon ions remains a signature for electron-induced segregation. 
  
We gratefully acknowledge support from Ch.\ Gnodtke during the early stage of this project.


\begin{thebibliography}{10}

\bibitem{hobo+08}
M. Hoener {\it et~al.}, J. Phys. B {\bf 41},  181001  (2008).

\bibitem{zich+11}
B. Ziaja {\it et~al.}, Phys. Rev. A {\bf 84},  033201  (2011).

\bibitem{suiw+12}
A. Sugishima {\it et~al.}, Phys. Rev. A {\bf 86},  033203  (2012).

\bibitem{misa+09}
A. Mikaberidze, U. Saalmann, and J.~M. Rost, Phys. Rev. Lett. {\bf 102},
  128102  (2009).

\bibitem{krfe+11}
S.~R. Krishnan {\it et~al.}, Phys. Rev. Lett. {\bf 107},  173402  (2011).

\bibitem{jhkr08}
J. Jha and M. Krishnamurthy, J. Phys. B {\bf 41},  041002  (2008).

\bibitem{spwe+12}
J.~C.~H. Spence, U. Weierstall, and H.~N. Chapman, Rep. Prog. Phys. {\bf 75},
  102601  (2012).

\bibitem{beti+04}
M. Bergh, N. T{\^\i}mneanu, and D. van~der Spoel, Phys. Rev. E {\bf 70},
  051904  (2004).

\bibitem{iwan+12}
B. Iwan {\it et~al.}, Phys. Rev. A {\bf 86},  033201  (2012).

\bibitem{gnsa+09}
C. Gnodtke, U. Saalmann, and J.~M. Rost, Phys. Rev. A {\bf 79},  041201\,(R)
  (2009).

\bibitem{anni+10}
A.~A. Andreev, P.~V. Nickles, and K.~Y. Platonov, Phys. Plasmas {\bf 17},
  023110  (2010).

\bibitem{gesa+07gnsa+12}
I. Georgescu, U. Saalmann, and J.~M. Rost, Phys. Rev. Lett. {\bf 99},  183002
  (2007);
C. Gnodtke, U. Saalmann, and J.-M. Rost, Phys. Rev. Lett. {\bf 108},  175003
  (2012).

\bibitem{thva09}
A. Thompson and D. Vaughan, X-Ray Data Booklet, 2009.

\bibitem{behu+98}
M. Berger {\it et~al.}, XCOM: Photon Cross Sections Data\-base, 1998.

\bibitem{koav11}
P. Koloren{\v c} and V. Averbukh, J. Chem. Phys. {\bf 135},  134314  (2011).

\bibitem{me85a}
W. Mehlhorn,  in {\em Atomic Inner-Shell Physics}, edited by B. Crasemann
  (Springer, New York, 1985), Chap.~Auger-Electron Spectrometry of Core Levels
  of Atoms, p.\ 119.

\bibitem{avsa+12}
V. Averbukh, U. Saalmann, and J.~M. Rost, Phys. Rev. A {\bf 85},  063405
  (2012).

\bibitem{inbu+12ingr+13}
L. Inhester, C.~F. Burmeister, G. Groenhof, and H. Grubm{\"u}ller, J. Chem.
  Phys. {\bf 136},  144304  (2012);
L. Inhester, G. Groenhof, and H. Grubm{\"u}ller, J. Chem. Phys. {\bf 138},
  164304  (2013).

\bibitem{sasi+06}
U. Saalmann, C. Siedschlag, and J.~M. Rost, J. Phys. B {\bf 39},  R\,39
  (2006).

\bibitem{gnsa+11}
C. Gnodtke, U. Saalmann, and J.-M. Rost, New J. Phys. {\bf 13},  013028
  (2011).

\end{thebibliography}
\end{document}